\documentclass[onecolumn,amsmath,amssymb]{qm2008_abs}
\usepackage{graphicx}
\usepackage{dcolumn}
\usepackage{bm}
\newcommand{\sqrts}{\sqrt{s}}

\providecommand{\PYTHIA}{{\sc pythia }}

\def\Et{\mbox{$E_{T}$}}
\def\ETa{E_{T,\,1}}
\def\ETb{E_{T,\,2}}

\begin{document}
\title{{\Large Low-x QCD studies with forward jets in p-p at 14 TeV}}
\bigskip
\bigskip
\author{\large Salim Cerci$^{1,}$\footnote{Supported by Fermilab US-CMS HCAL, and Turkish Atomic Energy Board (TAEK)} 
and David d'Enterria$^{2,}$\footnote{Support from 6th EU Framework Programme MEIF-CT-2005-025073 acknowledged.}\\ 
for the CMS collaboration}
\affiliation{$^1$ Cukurova University, Adana, Turkey}
\affiliation{$^2$ CERN, CH-1211 Geneva 23, Switzerland}

\bigskip
\bigskip

\begin{abstract}
\leftskip1.0cm
\rightskip1.0cm
The Large Hadron Collider will provide hadronic collisions at energies in the multi-TeV range,
never explored before. The parton fractional momenta probed at such energies
can be as low as $x \approx 2p_T/\sqrt{s}\, e^{-y}\approx 10^{-5}$  at large rapidities $y$, 
opening up attractive opportunities for low-$x$ QCD studies.
The combination of the CMS HF (3$<|\eta|<$5) and CASTOR (5.1$<|\eta|<$6.6) 
calorimeters allows one, in particular, to reconstruct very forward jets. 
We present generator-level studies of the CMS capabilities to measure the single inclusive 
forward jet spectrum and forward-backward (Mueller-Navelet) dijets in p-p collisions 
at 14 TeV. Both observables are sensitive to low-$x$ gluon densities and non-linear QCD evolution.
\end{abstract}
  
\maketitle


\section{Introduction}
\begin{figure}[htb]
\begin{center} 
\includegraphics[width=14.cm,height=6.cm, bb = 0 0 567 254,clip]{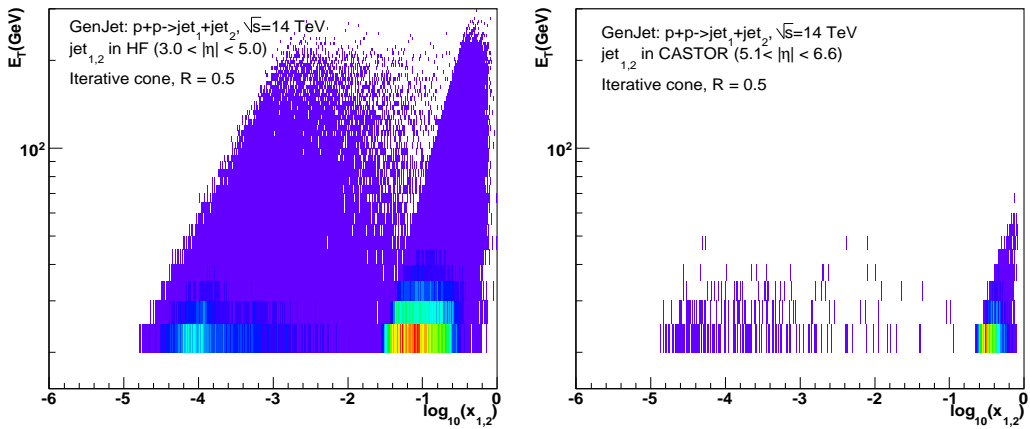}
\caption{log($x_{1,2}$) distribution of two partons producing at least one jet above 
$E_T$ = 20~GeV within HF ($3<|\eta|<5$, left) and CASTOR ($5.1<|\eta|<6.6$, right) 
in p-p collisions at $\sqrts$ = 14~TeV~\protect\cite{Fwd_LoI}.}
\label{fig:x1x2_hf_jets}
\end{center}
\end{figure}

The parton distribution functions (PDFs) in the proton have been studied in detail in  deep-inelastic-scattering
(DIS) $ep$ collisions at HERA~\cite{Klein:2008di}. 
For decreasing parton momentum fraction $x$~=~$p_{\mbox{\tiny{\it parton}}}/p_{\mbox{\tiny{\it hadron}}}$, 
the gluon density is observed to grow rapidly 
as $xg(x,Q^2)\propto x^{-\lambda(Q^2)}$, with $\lambda \approx$~0.1--0.3 rising
logarithmically with $Q^2$. As long as the densities are not too high, this growth is described 
by the Dokshitzer-Gribov-Lipatov-Altarelli-Parisi (DGLAP)~\cite{dglap} or by the 
Balitski-Fadin-Kuraev-Lipatov (BFKL)~\cite{bfkl} evolution equations which govern, respectively, 
parton radiation in $Q^2$ and $x$. 
Eventually, at high enough centre-of-mass energies (i.e. at very small $x$) the gluon density will 
be so large that non-linear (gluon-gluon fusion) effects will become important, saturating the 
growth of the parton densities~\cite{sat}. Studies of the high-energy (low-$x$) limit of QCD 
have attracted much theoretical interest in the last 10--15 years, in the 
context of DIS and of nucleus-nucleus collisions~\cite{cgc}.
Experimentally, direct information on the parton structure and evolution can be obtained  
in hadron-hadron collisions from the perturbative production of e.g. jets or prompt $\gamma$'s, 
which are directly coupled to the parton-parton scattering vertex. 
From leading-order (LO) kinematics, the rapidities and momentum fractions of the two colliding
partons are related via
\begin{equation}
x_{2} = (p_T/\sqrt{s})\cdot(e^{-y_1}+e^{-y_2}) \;\; \mbox{ and } \;\;
x_{1} = (p_T/\sqrt{s})\cdot(e^{y_1}+e^{y_2}).
\label{eq:x12}
\end{equation}
The {\it minimum} momentum fractions probed in a  $2\rightarrow 2$ process with a particle
of momentum $p_T$ produced at pseudo-rapidity $\eta$ are
\begin{equation}
x_{2}^{min} = \frac{x_T\,e^{-\eta}}{2-x_T\,e^{\eta}}\;\;,\;\;\;\; x_{1}^{min} = \frac{x_2\,x_T\,e^{\eta}}{2x_2-x_T\,e^{-\eta}}\;\;,
\mbox{ where } \;\; x_T=2p_T/\sqrt{s}\,,
\label{eq:x2_min}
\end{equation}
i.e. $x_2^{min}$ decreases by a factor of $\sim$10 every 2 units of rapidity.
From Eq.~(\ref{eq:x2_min}), it follows that the measurement of jets with transverse energy $E_{T}\approx$~20~GeV 
in the CMS forward calorimeters (HF, 3$<|\eta|<$5 and CASTOR, 5.1$<|\eta|<$6.6) will allow one 
to probe $x$ values as low as $x_{2}\approx 10^{-5}$. Figure~\ref{fig:x1x2_hf_jets} shows 
the actual log($x_{1,2}$) distribution for two-parton scattering in p-p collisions at 14~TeV producing at least 
one jet above 20 GeV in the HF and CASTOR acceptances. 
We present here generator-level studies of two forward-jet measurements in CMS sensitive to small-$x$ QCD~\cite{Fwd_LoI}:
\begin{enumerate}
\item single inclusive jet cross section in HF at moderately high virtualities ($E_{T}\approx$ ~20 -- 100~GeV),
\item differential cross sections and azimuthal (de)correlation of ``Mueller-Navelet'' (MN)~\cite{mueller_navelet} 
dijet events, characterized by jets with 
similar $E_T$ separated by a large rapidity interval ($\Delta\eta\approx$~10).
\end{enumerate}
The first measurement is sensitive to the low-$x_2$ (and high-$x_1$) proton PDFs,
whereas the second one yields information on  BFKL-~\cite{mueller_navelet,orr,DelDuca93,sabiovera} 
and saturation-~\cite{marquet,iancu08} type dynamics.

\section{Experimental Setup}

\begin{figure}[htb]
\begin{center}
\includegraphics[width=13.cm,height=5.cm]{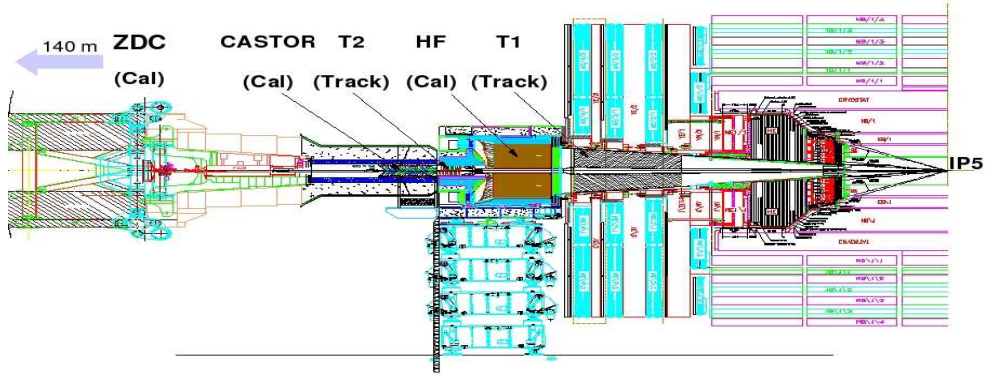}
\caption{Layout of the detectors in the CMS forward region used for the low-$x$ QCD studies~\protect\cite{Fwd_LoI}.
\label{fig:forw_CMS}}
\end{center}
\end{figure}

The combination of HF, TOTEM, CASTOR and ZDC (Fig.~\ref{fig:forw_CMS}) makes of CMS 
the largest acceptance experiment ever built at a collider. Very forward jets can be identified 
using the HF~\cite{hf} and CASTOR~\cite{castor} calorimeters. The HF, located 11.2 m away on both sides of the 
interaction point (IP), is a steel plus quartz-fiber \v{C}erenkov calorimeter segmented into 1200 towers of 
$\Delta\eta\times\Delta\phi\sim$~0.175$\times$0.175. It has $10 \lambda_I$ interaction lengths
and is sensitive to deposited electromagnetic (EM) and hadronic (HAD) energy.
CASTOR is an azimuthally symmetric EM/HAD calorimeter placed at 14.37 m from
the IP, covering 5.1$<|\eta|<$6.6. The calorimeter is a \v{C}erenkov-light device,
with successive layers of tungsten absorber and quartz plates as active medium
arranged in 2 EM (10 HAD) sections of about 22$X_0$ (10.3$\lambda_I$) radiation (interaction) lengths.

\section{Forward jets reconstruction in HF}
\label{sec:fwd_jets}

Jets in CMS are reconstructed at the generator- and calorimeter-level using 3 different jet algorithms~\cite{CMS_AN-2008/002}:
iterative cone~\cite{ptdr1} with radius of ${\cal R}=0.5$ in $(\eta,\phi)$, SISCone
~\cite{Salam_SISCone} (${\cal R}=0.5$), and the Fast-$k_{T}$~\cite{Cacciari_FastKt} 
($E_{seed}$~=~3~GeV and $E_{thres}$ = 20~GeV). 
The Monte Carlo samples used in this analysis were part of the official CMS QCD-jets simulation using 
PYTHIA~\cite{pythia6.4} in seven 1M-events $\hat{p}_{T}$ bins 
across the $E_T$~=~15--230 GeV range. Events are selected where at least one jet above 20 GeV 
falls in the forward HF acceptance.
The matching radius between generated and reconstructed jets, for reconstruction performance studies, 
is set at $\Delta$R = 0.2. The $E_{T}$ resolutions for the three different algorithms 
are very similar: $\sim$18\% at $E_{T}\sim$20 GeV decreasing to  $\sim$12\% for $E_{T}\gtrsim$100 GeV
(Fig.~\ref{fig:dsigmadET_fwd_jets_and_ETresolution}, left). The position ($\eta$, $\phi$) 
resolutions (not shown here) for jets in HF are also very good: $\sigma_{\phi,\eta}$~=~0.045 at $E_T$~=~20 GeV, 
improving to $\sigma_{\phi,\eta}\sim$~0.02 above 100 GeV. The forward jet energy scale uncertainty is, 
however, expected to be relatively large (in the range 10\%--3\% in the same $E_T$ range)~\cite{ptdr1}.


\section{Single inclusive forward jet measurement}

\begin{figure}[htb]
\begin{center}
\includegraphics[width=0.49\textwidth,height=6.cm]{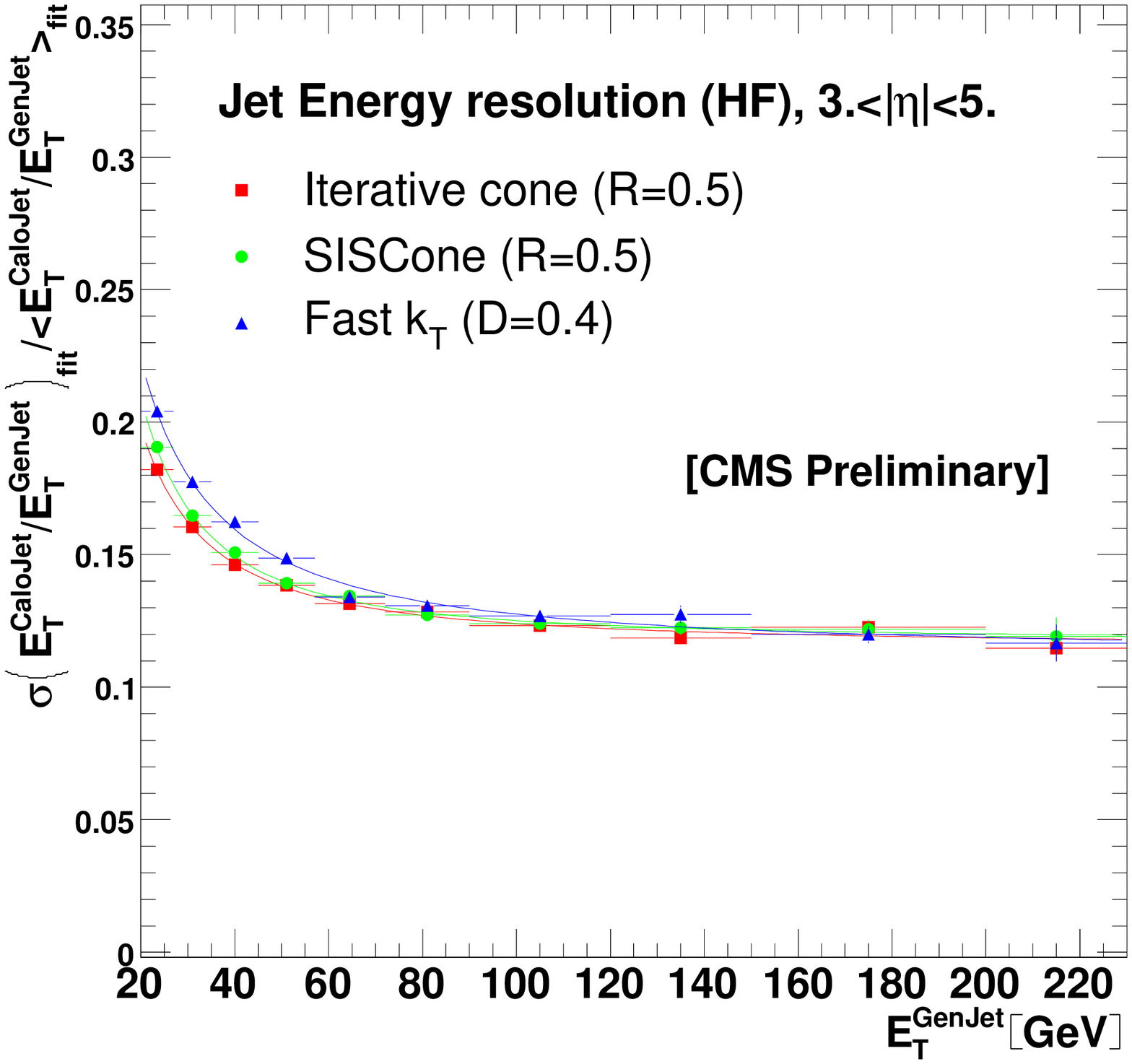}
\includegraphics[width=0.40\textwidth,height=6.cm]{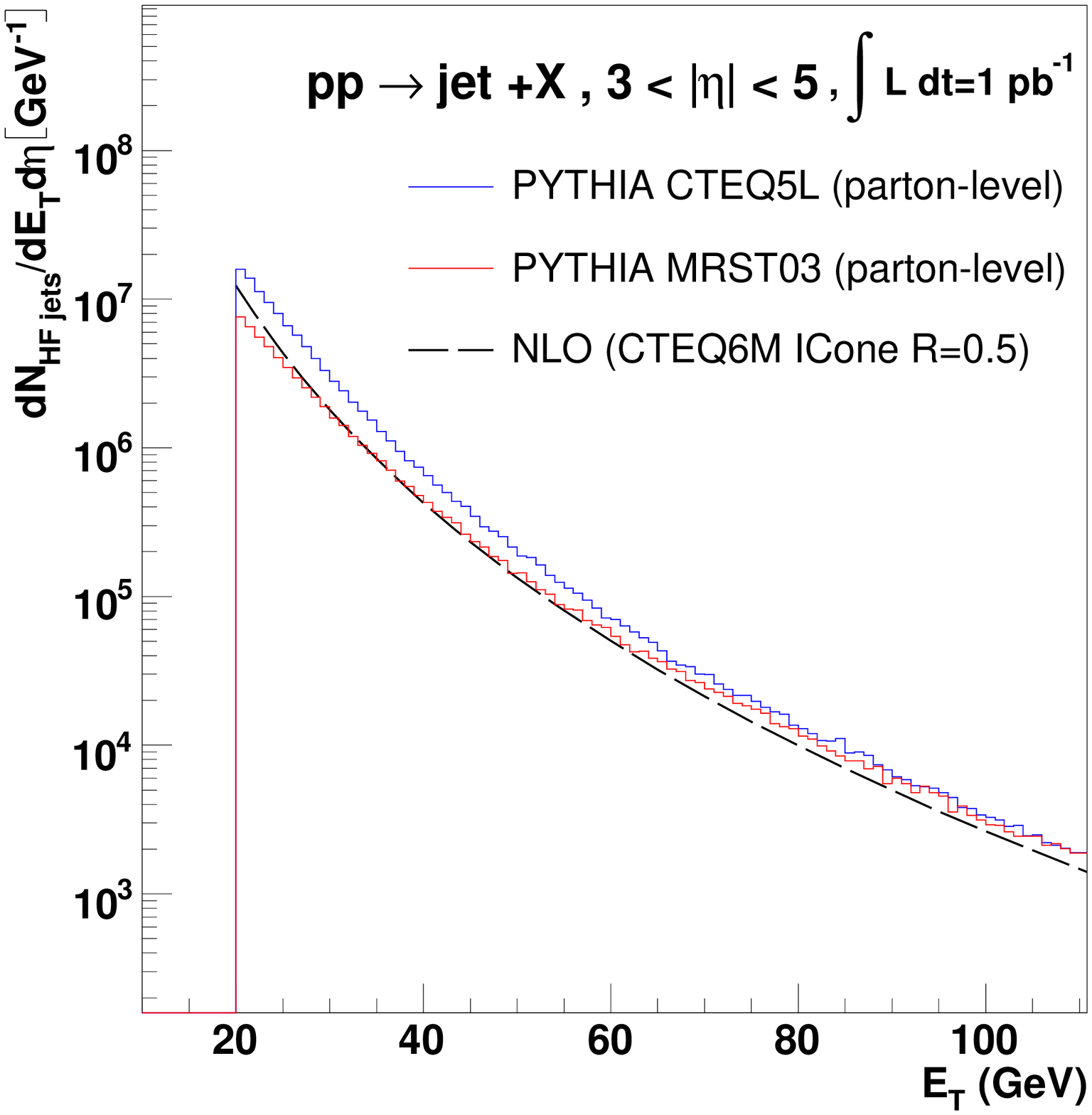}
\caption{Left: Energy resolution for jets measured in HF as a function of $E_{T}$~\protect\cite{CMS_AN-2008/002}. 
Right: Single inclusive jet yields in HF expected in (1~fb$^{-1}$) p-p collisions at 14~TeV obtained with \PYTHIA 6.403 
at the parton-level with CTEQ5L and MRST03 PDFs (histograms) compared to a NLO jet calculation using CTEQ6M~\protect\cite{vogelsang}.
\label{fig:dsigmadET_fwd_jets_and_ETresolution}}
\end{center}
\end{figure}

Figure~\ref{fig:dsigmadET_fwd_jets_and_ETresolution} right shows the single jet spectrum expected 
in both HFs for 1~pb$^{-1}$ integrated luminosity obtained at the parton-level from \PYTHIA 
for two different PDF sets (CTEQ5L and MRST03) compared to a NLO calculation (CTEQ6M, 
${\cal R}=0.5$, scales $\mu$ = 0.5$\Et$--2$\Et$)~\cite{vogelsang}. 
The single jet spectra obtained for different PDFs are similar at high $E_{T}$, while differences
as large as $\mathcal{O}(60\%)$ appear below $\sim$60~GeV. The measurement of low-$E_T$ forward jets in HF 
seems in principle feasible: the statistical errors are negligible and the HF energy {\it resolution} is very good 
(Fig.~\ref{fig:dsigmadET_fwd_jets_and_ETresolution}, left). Yet, in the ``interesting'' low-$E_T$ range, 
the main experimental issue will be the control of the jet-energy {\it scale} whose uncertainty 
propagates into up to $\pm$40\% differences in the final jet yield. Use of this measurement 
to constrain the proton PDFs in the low-$x$ range will thus require careful studies of the HF jet 
calibration.


\section{Mueller-Navelet (MN) dijets measurement}

\begin{figure}[htb]
\begin{center}
\includegraphics[width=0.49\textwidth,height=6.cm]{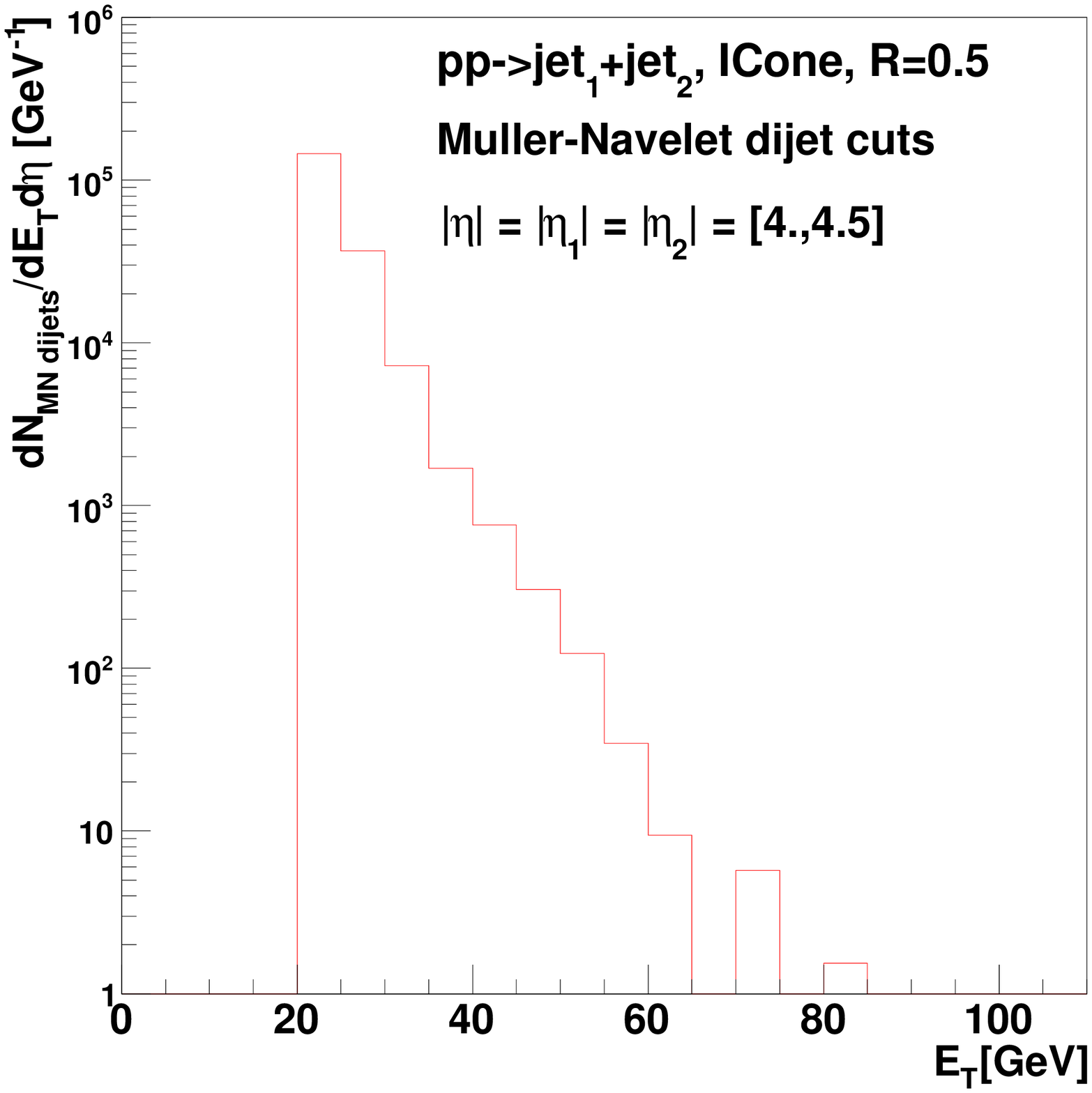}
\includegraphics[width=0.49\textwidth,height=6.cm]{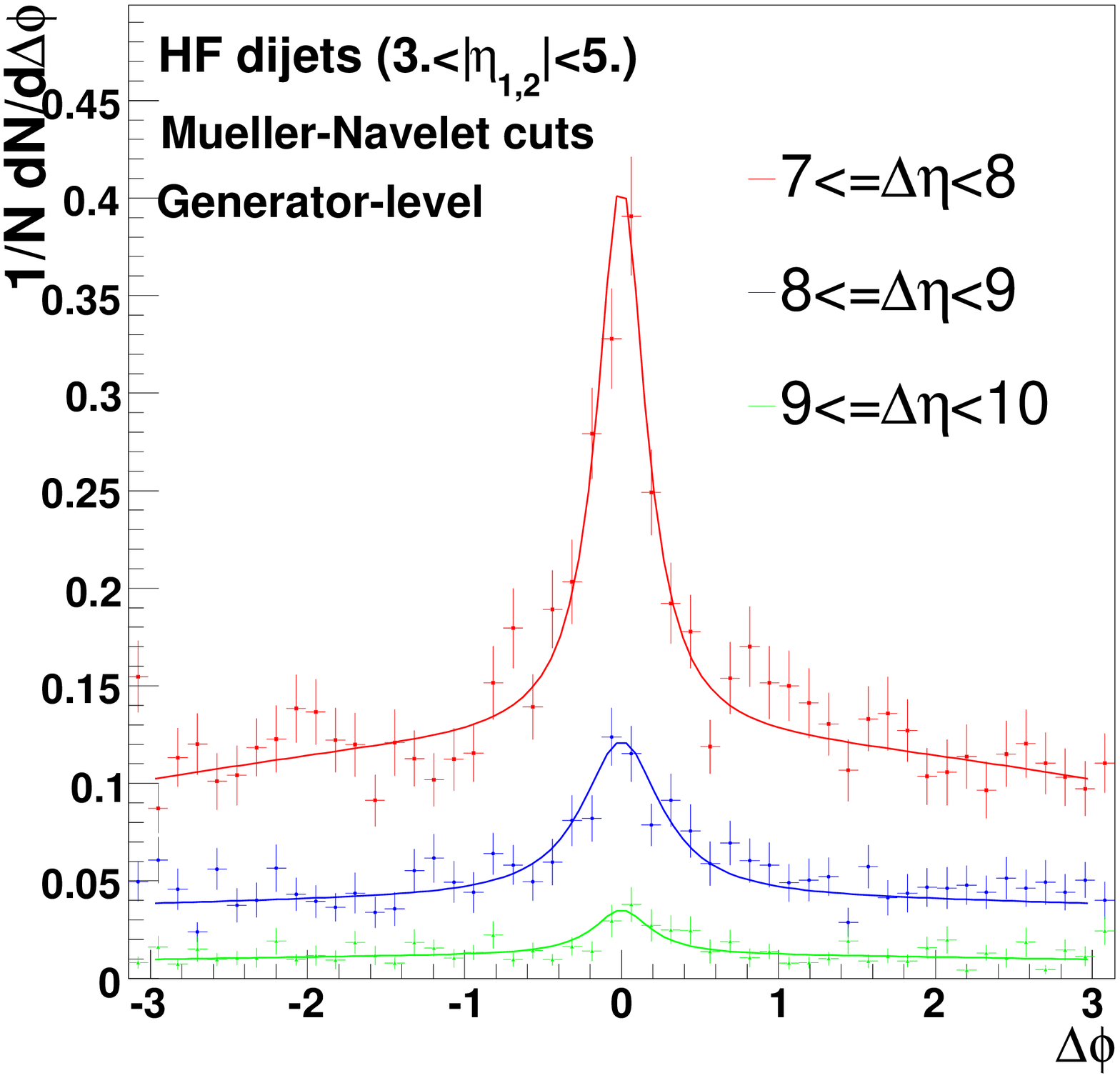}
\caption{Dijet events passing the Mueller-Navelet cuts described in the text. 
Left: Expected yields (in 1~pb$^{-1}$) for a separation $\Delta\eta \approx$~8~\protect\cite{Fwd_LoI}. 
Right: $\Delta\phi$ distributions for jet separations $\Delta\eta$~=~7.5, 8.5 and 9.5. 
\label{fig:muller_navelet_jets_dphi}}
\end{center}
\end{figure}

Inclusive dijet production at large pseudorapidity intervals in high-energy hadron-hadron collisions 
has been since long considered an excellent testing ground for BFKL~\cite{mueller_navelet,DelDuca93,orr,sabiovera} 
and also for saturation~\cite{marquet,iancu08} QCD evolutions.
Both colliding partons in the MN kinematics are large-$x$ valence quarks ($x_{1,2}\approx$~0.2), 
which produce two jets with transverse energies $E_{T\,,i}$ with a large rapidity interval between them:
\begin{equation}
Y=\log(x_1\,x_2\, s/(Q_1\,Q_2))\;,
\label{eq:Y}
\end{equation}
where $Q_{i}\approx E_{T\,,i}$ are the corresponding parton virtualities.
The presence of a large rapidity separation ($Y=\Delta\eta$) between jets enhances the available phase-space 
in longitudinal momentum for extra BFKL-type radiation.
In CMS, jet rapidity separations as large as $\Delta\eta\approx$~12 are accessible combining
both HF and CASTOR opposite hemispheres. As a proof of principle, we have reanalyzed the \PYTHIA
jet samples discussed in the previous Section, and selected events which satisfy the following 
Mueller-Navelet-type selection cuts:
\begin{itemize}
\item $E_{T,i} > 20$~GeV
\item $|\ETa - \ETb| < 5$~GeV (similar virtuality, $Q \approx \sqrt{\ETa \cdot\ETb}$, to minimise DGLAP evolution)
\item $3 <|\eta_{1,2}|< 5$ (both jets in HF) 
\item $\eta_1 \cdot \eta_2 < 0$  (each jet in a different HF)
\item $||\eta_1| - |\eta_2|| < 0.5$ (almost back-to-back in pseudo-rapidity)
\end{itemize}
The data have been divided into 4 equidistant HF pseudorapidity bins 
and the dijet cross sections in each $\eta$ bin computed as $d^2\sigma/d\eta dQ = N_{jets}/(\Delta\eta \Delta Q \int\mathcal{L}\mbox{dt})$,
where $N$ is the observed number of  jets in the bin and 1~pb$^{-1}$ the assumed integrated luminosity.
The left plot in Fig.~\ref{fig:muller_navelet_jets_dphi} shows the expected dijet yields passing 
the MN kinematics cuts as a function of $Q$ for the pseudo-rapidity separation 
$\Delta\eta\approx$~8. 
The obtained MN dijet statistics appears large enough to carry out detailed studies of the $\Delta\eta$
dependence, that would e.g. provide evidence for a possible Mueller-Navelet ``geometric scaling'' behaviour~\cite{iancu08}.
An enhanced azimuthal decorrelation for increasing rapidity separation between the Mueller-Navelet jets 
is the classical ``smoking-gun'' of BFKL radiation~\cite{orr,DelDuca93,sabiovera}. The generator-level $\Delta\phi$ 
jet distributions are plotted in the right plot of Fig.~\ref{fig:muller_navelet_jets_dphi} for $\Delta\eta$~=~7.5, 8.5 and 9.5. 
The peak at $\Delta\phi$ = $(\phi_{1}-\phi_{2})-\pi$~=~0 indicates that the two jets are highly correlated 
with each other. As the $\Delta\eta$ between the two jets increases, the peaks diminish and the distributions
get increasingly larger, signaling a loss in correlation. Since \PYTHIA is a leading-order generator without 
any BFKL (or saturation) effect, the observed azimuthal decorrelation is just due to parton shower effects
and initial- or final-state radiation. Such a result provides, thus, a baseline of the minimal decorrelation expected
in non-BFKL scenarios. Detailed simulation studies are ongoing to test the sensitivity of such forward jet 
measurements to signal (or not) the presence of ``genuine'' low-$x$ decorrelations.\\

\newpage


\end{document}